\documentclass[twocolumn,iop,numberedappendix,appendixfloats]{openjournal}

\usepackage[utf8]{inputenc}
\usepackage[british]{babel}

\usepackage{afterpage}
\usepackage{rotating}

\usepackage{amsmath}	
 \usepackage{amssymb}	
\usepackage{subcaption}  
\usepackage{caption}

\usepackage{xcolor}
\definecolor{maroon}{cmyk}{0, 0.87, 0.68, 0.32}
\definecolor{halfgray}{gray}{0.55}
\definecolor{ipython_frame}{RGB}{207, 207, 207}
\definecolor{ipython_bg}{RGB}{247, 247, 247}
\definecolor{ipython_red}{RGB}{186, 33, 33}
\definecolor{ipython_green}{RGB}{0, 128, 0}
\definecolor{ipython_cyan}{RGB}{64, 128, 128}
\definecolor{ipython_purple}{RGB}{170, 34, 255}
\usepackage{float}
\usepackage{fontawesome}
\usepackage[colorlinks,allcolors=blue]{hyperref}
\usepackage{url}

\usepackage{listings}

\lstdefinelanguage{iPython}{
    morekeywords={access,and,break,class,continue,def,del,elif,else,except,exec,finally,for,from,global,if,import,in,is,lambda,not,or,pass,print,raise,return,try,while},%
    morekeywords=[2]{abs,all,any,basestring,bin,bool,bytearray,callable,chr,classmethod,cmp,compile,complex,delattr,dict,dir,divmod,enumerate,eval,execfile,file,filter,float,format,frozenset,getattr,globals,hasattr,hash,help,hex,id,input,int,isinstance,issubclass,iter,len,list,locals,long,map,max,memoryview,min,next,object,oct,open,ord,pow,property,range,raw_input,reduce,reload,repr,reversed,round,set,setattr,slice,sorted,staticmethod,str,sum,super,tuple,type,unichr,unicode,vars,xrange,zip,apply,buffer,coerce,intern},%
    sensitive=true,%
    morecomment=[l]\#,%
    morestring=[b]',%
    morestring=[b]",%
    morestring=[s]{'''}{'''},
    morestring=[s]{"""}{"""},
    morestring=[s]{r'}{'},
    morestring=[s]{r"}{"},%
    morestring=[s]{r'''}{'''},%
    morestring=[s]{r"""}{"""},%
    morestring=[s]{u'}{'},
    morestring=[s]{u"}{"},%
    morestring=[s]{u'''}{'''},%
    morestring=[s]{u"""}{"""},%
    literate=
    {á}{{\'a}}1 {é}{{\'e}}1 {í}{{\'i}}1 {ó}{{\'o}}1 {ú}{{\'u}}1
    {Á}{{\'A}}1 {É}{{\'E}}1 {Í}{{\'I}}1 {Ó}{{\'O}}1 {Ú}{{\'U}}1
    {à}{{\`a}}1 {è}{{\`e}}1 {ì}{{\`i}}1 {ò}{{\`o}}1 {ù}{{\`u}}1
    {À}{{\`A}}1 {È}{{\'E}}1 {Ì}{{\`I}}1 {Ò}{{\`O}}1 {Ù}{{\`U}}1
    {ä}{{\"a}}1 {ë}{{\"e}}1 {ï}{{\"i}}1 {ö}{{\"o}}1 {ü}{{\"u}}1
    {Ä}{{\"A}}1 {Ë}{{\"E}}1 {Ï}{{\"I}}1 {Ö}{{\"O}}1 {Ü}{{\"U}}1
    {â}{{\^a}}1 {ê}{{\^e}}1 {î}{{\^i}}1 {ô}{{\^o}}1 {û}{{\^u}}1
    {Â}{{\^A}}1 {Ê}{{\^E}}1 {Î}{{\^I}}1 {Ô}{{\^O}}1 {Û}{{\^U}}1
    {œ}{{\oe}}1 {Œ}{{\OE}}1 {æ}{{\ae}}1 {Æ}{{\AE}}1 {ß}{{\ss}}1
    {ç}{{\c c}}1 {Ç}{{\c C}}1 {ø}{{\o}}1 {å}{{\r a}}1 {Å}{{\r A}}1
    {€}{{\EUR}}1 {£}{{\pounds}}1
    {^}{{{\color{ipython_purple}\^{}}}}1
    {=}{{{\color{ipython_purple}=}}}1
    {+}{{{\color{ipython_purple}+}}}1
    {-}{{{\color{ipython_purple}-}}}1
    {*}{{{\color{ipython_purple}$^\ast$}}}1
    {/}{{{\color{ipython_purple}/}}}1
    {+=}{{{+=}}}1
    {-=}{{{-=}}}1
    {*=}{{{$^\ast$=}}}1
    {/=}{{{/=}}}1,
    literate=
    *{-}{{{\color{ipython_purple}-}}}1
     {?}{{{\color{ipython_purple}?}}}1,
    identifierstyle=\color{black}\ttfamily,
    commentstyle=\color{ipython_cyan}\ttfamily,
    stringstyle=\color{ipython_red}\ttfamily,
    keepspaces=true,
    showspaces=false,
    showstringspaces=false,
    rulecolor=\color{ipython_frame},
    frameround={t}{t}{t}{t},
    numbers=none,
    numberstyle=\tiny\color{halfgray},
    backgroundcolor=\color{ipython_bg},
    basicstyle=\ttfamily\footnotesize,
    columns=fullflexible,
    keywordstyle=\color{ipython_green}\ttfamily,
}

\usepackage{array,multirow,graphicx}

\newcommand{\be}{\begin{equation}}
\newcommand{\ee}{\end{equation}}
\newcommand{\bea}{\begin{eqnarray}}
\newcommand{\eea}{\end{eqnarray}}

\newcommand{\lcdm}{$\Lambda$CDM}

\newcommand{\Omm}{\Omega_{\rm m}}

\newcommand{\Ads}{A_{\rm ds}}
\newcommand{\ns}{n_{\rm s}}
\newcommand{\Adsp}{A_{\rm ds, piv}}

\newcommand{\refr}[1]{\textcolor{black}{#1}}

\begin{document}

\journalinfo{The Open Journal of Astrophysics}
\submitted{submitted XXX; accepted YYY}

\title{Evolving and interacting dark energy: photometric and spectroscopic synergy with DES Y3 and DESI DR2}

\shorttitle{CPL \& DS: DES+DESI}
\shortauthors{M. Tsedrik  \& B. Bose}

\author{M. Tsedrik$^{1,2}$  \thanks{E-mail: \href{maria.tsedrik@ed.ac.uk}{maria.tsedrik@ed.ac.uk}}}
\author{B. Bose$^{1,2}$} 
\affiliation{$^{1} $Institute for Astronomy, University of Edinburgh, Royal Observatory, Blackford Hill, Edinburgh, EH9 3HJ, UK} 
\affiliation{$^{2}$Higgs Centre for Theoretical Physics, School of Physics and Astronomy, Edinburgh, EH9 3FD, UK}

\begin{abstract}
We investigate the Dark Scattering (DS) interacting dark energy scenario, characterised by pure momentum exchange between dark matter and dark energy, \refr{combined with} a time-dependent equation-of-state \refr{for dark energy} described by the Chevallier–Polarski–Linder (CPL) parametrisation. This class of models is weakly constrained by CMB observations and can exhibit distinctive late-time suppression of structure growth. We derive constraints on cosmological, DS, and CPL parameters using three two-point correlation functions from the Dark Energy Survey Year 3 data, combined with baryon acoustic oscillation measurements from DESI, Type Ia supernovae from DES Year 5, and CMB data from Planck. We find the dark-sector interaction parameter $\Ads$ to be consistent with zero for all data combinations, and that CPL provides a statistically preferred fit over DS for the selected probes. From the full data combination we obtain $w_0=-0.76\pm0.06, \, w_a=-0.77^{+0.23}_{-0.20}$ for CPL, and $w_0=-0.79^{+0.05}_{-0.06}, \, w_a=-0.56^{+0.24}_{-0.15}, \, (\Ads=9.8^{+2.8}_{-9.5}\,\mathrm{bn/GeV} )$ for DS. The inclusion of DES photometric information improves the Figure-of-Merit on $(w_0,w_a)$ by \refr{$\sim$12\%} for CPL and \refr{$\sim$25\%} for DS relative to DESI+SN+CMB alone. We find no evidence for an $S_8$ discrepancy \refr{between the low-$z$ and high-$z$ measurements} in either model. These results provide the most stringent pre-Euclid constraints on DS from a combined photometric and spectroscopic analysis.
\end{abstract}


\maketitle

\section{Introduction}
\label{sec:introduction}

The new generation of cosmological surveys such as DESI~\footnote{\url{https://www.desi.lbl.gov}}, Euclid~\footnote{\url{https://www.euclid-ec.org}}, and Rubin~\footnote{\url{https://rubinobservatory.org}} will deliver Large-Scale Structure (LSS) measurements of unprecedented precision, offering new insights into the nature of the dark sector -- dark energy, dark matter and their interactions. The quality of these data will enable decisive tests of the current concordance cosmological model \lcdm, and will elucidate growing tensions within it~\citep{CosmoVerse:2025txj}. These tensions include the recent measurement by the DESI experiment that strongly suggests a time-evolving dark energy component~\citep{Collaboration-2025-DESIDR2Resultsb} and the recently softened $S_8$ tension~\citep{Wright:2025xka}, a discrepancy between the amplitude of matter fluctuations inferred from LSS observations~\citep{kids2021, des2022} and the value predicted from Cosmic Microwave Background (CMB) measurements extrapolated to late times within \lcdm~\citep{planck2018cosmo}.

Among the proposed extensions to \lcdm{} capable of addressing these emerging tensions are interacting dark energy models~\citep{Pourtsidou:2013nha}. These models are also of high theoretical interest as any detection of an interaction would contain valuable information about the microphysics of the dark sector. In particular, pure momentum exchange, or Type III models, offer distinct phenomenology, allowing for the suppression of structure growth while leaving the background unchanged. This was favorable in light of the $S_8$ tension, but also provided a more fundamental theoretical model to fit the DESI measurements. A phenomenological representative of these Type III models is the well studied Dark Scattering (DS) model~\citep{Simpson:2010vh}, in which dark energy and cold dark matter interact through a pure momentum-exchange mechanism governed by a coupling parameter controlling their cross-section. This coupling modifies the growth of cosmic structure while leaving the background expansion largely unaffected unlike energy-exchange models. The DS framework has been extensively developed through theoretical work, simulations, and data analyses~\citep{Baldi:2014ica, Baldi:2016zom, Pourtsidou:2016ico, Kumar:2017bpv, Bose:2017jjx, Bose:2018zpk, Carrilho:2021hly, Carrilho:2021otr, ManciniSpurio:2021jvx, Linton:2021cgd, Tsedrik:2022cri, Carrilho:2022mon, Carrion:2024itc, Tsedrik:2025cwc}.

\begin{figure*}
\centering
    \includegraphics[width=0.9\textwidth]{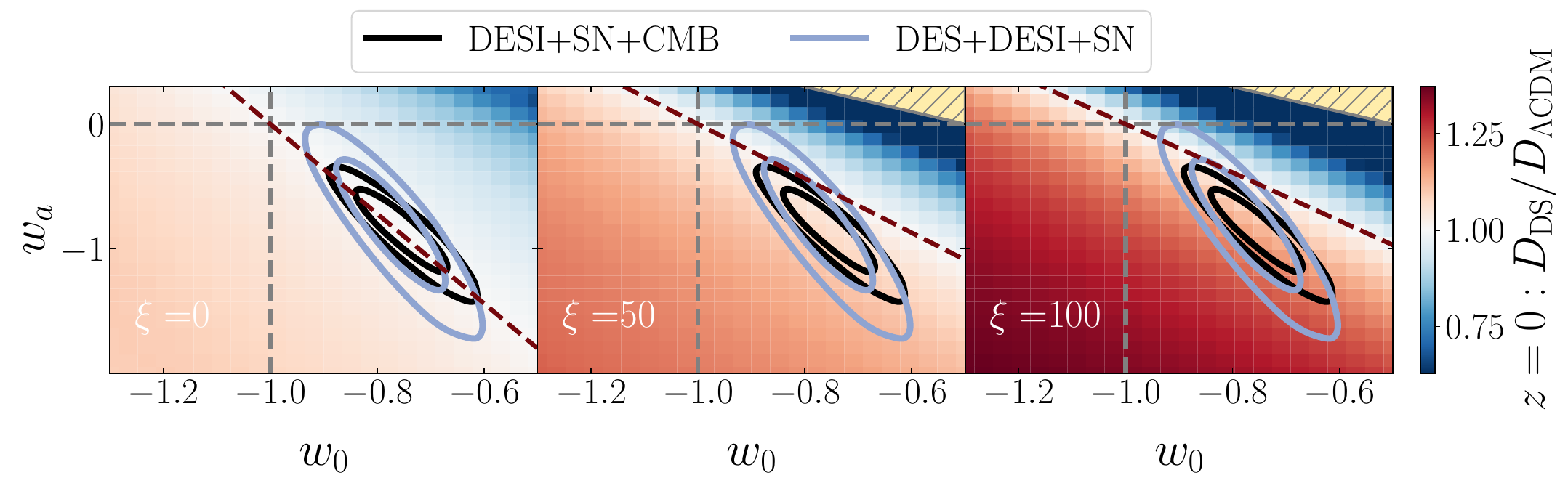}
    \caption{Linear growth factor in DS with respect to \lcdm{} today, overlayed with \href{https://data.desi.lbl.gov/public/papers/y3/bao-cosmo-params/README.html}{official} DESI results for evolving dark energy without interaction (specific combinations of cosmological probes are shown in the legend). The reddish region on the heatmap denotes enhancement of structure growth with respect to \lcdm{}, while the blueish region denotes suppression. Grey dashed lines mark the \lcdm-limit for the dark energy equation-of-state. Dark red dashed line denotes the suppression/enhancement transition, i.e. $D_\mathrm{DS} \approx D_{\Lambda \mathrm{CDM}}$, for different values of the interaction strength $\xi$ (in units of [bn/GeV]).}
    \label{fig:linear-growth}
\end{figure*}

In this study, we investigate an extended DS scenario in which the dark energy equation-of-state evolves with time following the Chevallier–Polarski–Linder (CPL) parametrisation~\citep{Chevallier:2000qy,Linder:2002et}. We constrain this model using the Dark Energy Survey's~\footnote{\url{https://www.darkenergysurvey.org/}} year 3 (DES Y3) measurements of three two-point correlation functions -- cosmic shear, galaxy clustering, and the shear–position cross-correlation -- combined with the DR2 DESI BAO data, which has independently shown indications of time-evolving dark energy. This joint analysis leverages the complementarity between photometric and spectroscopic probes to assess both the interaction strength and the evolution of the dark energy component. To model nonlinear structure formation in the DS model with a CPL parametrised equation-of-state, we make use of the halo model reaction formalism~\citep{Cataneo:2018otr} and its associated code {\tt ReACT}~\citep[\href{https://github.com/nebblu/ACTio-ReACTio}{\faicon{github}}]{Bose:2020otr}. We construct a neural network emulator using the {\tt Cosmopower}~\citep[\href{https://alessiospuriomancini.github.io/cosmopower/}{\faicon{github}}]{SpurioMancini:2021ppk} platform, extending the work of \cite{Carrion:2024itc}. We make this emulator public on the {\tt ReACT}-based emulator repository (\href{https://github.com/nebblu/MGEmus}{\faicon{github}}). 

This paper is organised as follows: in \autoref{sec:model} we introduce the DS model and its background, linear and nonlinear structure formation predictions; in \autoref{sec:data_and_methods} we introduce the emulator and describe the data we use; in \autoref{sec:results} we present our results; we conclude and summarize in \autoref{sec:conclusions}.

\section{Model}
\label{sec:model}

The Dark Scattering (DS) model is an interacting dark energy scenario in which dark energy and cold dark matter exchange momentum through elastic scattering \citep{Simpson:2010vh, Pourtsidou:2013nha, Skordis:2015yra}. Since no energy is transferred between the two components, the background expansion history is not directly altered by the interaction. This form of coupling also has only a mild effect on the CMB, remaining in excellent agreement with Planck measurements \citep{Pourtsidou:2016ico}.

Compared to \lcdm{}, the DS framework introduces two additional parameters: the dark energy equation-of-state $w$ and the interaction strength $\xi$, the latter defined as the ratio of the scattering cross section to the dark matter particle mass. In this work, we generalise the equation-of-state to evolve in time according to the CPL parametrisation~\citep{Chevallier:2000qy,Linder:2002et}: 
\begin{equation}
    w(a) = w_0 + w_a(1-a)\, , 
    \label{eq:cpl}
\end{equation}
where $a$ is the scale factor and $\{w_0,w_a\}$ are free parameters. This paramtrisation allows us to explore scenarios in which dark energy dynamics and momentum exchange jointly affect structure growth. The background expansion in CPL is given by 
\begin{equation}
    H^2(a) = H_0^2 \left[\Omega_{\rm m,0} a^{-3} + \Omega_{\rm DE}\right] \, , 
    \label{eq:cpl_background}
\end{equation}
with
\begin{equation}
    \Omega_{\rm DE}(a) = \Omega_{{\rm DE},0}e^{\int^a_13[1+w(\tilde{a})]\tilde{a}\,d\tilde{a}} \, . 
\end{equation}
$H_0$ is the Hubble constant and $\Omega_{{\rm DE},0}$ and $\Omega_{\rm m,0}$ are the present day density parameters of dark energy and total matter respectively, with $\Omega_{{\rm DE},0} = 1-\Omega_{\rm m,0}$ for a flat Universe which we consider in this paper. 

The linear perturbations in DS are given by the Continuity and Euler equations~\citep{Simpson:2010vh,Baldi:2014ica,Bose:2017jjx}: 
\begin{eqnarray}
&&a \delta' +\theta = 0 \, ,
\label{eq:Perturb1}\\
&& a \theta' +
\left(2+A+\frac{a  H'}{H}\right)\theta
-\left(\frac{k}{a\,H}\right)^2\,\Phi= 0 \, ,
\label{eq:Perturb2}
\end{eqnarray}
where a prime denotes a scale factor derivative, $\delta$ and $\theta=\frac{\nabla\cdot \mathbf{v}}{a H}$ are the linear total density and velocity divergence perturbations, and $\Phi$ is the Newtonian gravitational potential. DS introduces an additional term into the Euler equation that controls a frictional force that slows or enhances the evolution of perturbations in LSS depending on the sign of $w$
\begin{equation}
A(a) \equiv [1+w(a)]\frac{H_0^2}{H(a)}\frac{3\xi}{8\pi G}\Omega_{\rm DE}(a) \, ,
\label{eq:Ads}
\end{equation}
where $\xi$ is a strictly positive parameter, \refr{defined as the CDM-dark energy scattering cross section per dark matter particle mass~\citep{Baldi:2014ica}}. It governs the interaction strength and has units of [bn/GeV]. The \lcdm{}-limit is given by $\xi=0$ and $w_0=-1$, $w_a=0$. 
In the linearised growth \autoref{eq:Perturb2} we evolve dark matter and baryons (total matter) as a single species for simplicity. In reality, interaction with dark energy affects dark matter but not baryons in the DS scenario. To take this non-universality of coupling into account, we actually suppress the interaction parameter $\xi$ by the dark matter fraction of the total matter, $f_c = \rho_{\rm c}/\rho_{\rm m}$ following~\citet{Carrilho:2021otr}:
\begin{equation}
    \bar{\xi} = \frac{f_c}{1+A(a=1) (1-f_c)}\, \xi \, .
\end{equation}
We have verified that the linear growth agrees with the modified version of \texttt{CLASS}~(\href{https://github.com/PedroCarrilho/class_public/tree/IDE_DS}{\faicon{github}}) within 1-2\% for values of DS parameters within our prior range.

The growth equation can be solved to find the linear growth factor, $D$, and the growth rate $f \equiv {\rm d} \ln D / {\rm d} \ln a$. For the linear growth factor to have a growing solution the frictional force must vanish at high redshifts, i.e.~$\lim_{a\rightarrow0}A(a)\rightarrow 0$. Hence, from \autoref{eq:Ads} we find the condition $w_0+w_a<-1/2$, which is stronger than the matter-domination condition \refr{$w_0+w_a<0$}. 

In \autoref{fig:linear-growth} we show for varying values of the interaction parameter the linear growth factor in DS with respect to \lcdm{} at $z=0$. We notice that the suppression/enhancement transition depends on the value of $\xi$: the suppression of structure growth is observed for  approximately 
$w_a\gtrsim -3.6\,(1+w_0)$ with $\xi=0$, $w_a\gtrsim -2.18 \, (1+w_0)$ with $\xi=50$,  and $w_a\gtrsim -1.94 \, (1+w_0)$ with $\xi=100$. We over-plot official DESI DR2 chains \citep{DESI:2025zgx} in the CPL-parametrisation of dark energy without interaction. The tightest constraints are obtained from the joint analysis of BAO with CMB and SN -- this combination of $(w_0,w_a)$ values in DS leads to enhancement of structure growth with respect to $\Lambda$CDM for all non-zero interaction values (the larger $\xi$, the stronger the enhancement). Substituting CMB with photometric information from DES Y3, while being less constraining, does not completely exclude the $(w_0,w_a)$ parameter space that yields suppression of structure growth with respect to \lcdm{}. \refr{This statement appears to hold true at $z>0$: while the overall suppression/enhancement distribution in the $w_0-w_a$ plane remains similar to the one in \autoref{fig:linear-growth}, the line signifying equality with $\Lambda$CDM growth slowly rotates anti-clockwise with increasing redshift. For instance, at $z=2$  the suppression of structure growth is observed for  approximately 
$w_a\gtrsim -2\,(1+w_0)$ with $\xi=0$, $w_a\gtrsim -1.5 \, (1+w_0)$ with $\xi=50$,  and $w_a\gtrsim -1.2 \, (1+w_0)$ with $\xi=100$.}

In previous analyses \citep{Carrilho:2022mon, Carrion:2024itc, Tsedrik:2025cwc}, instead of constraining $w_0$ and $\xi$, a new parameter $A_{\rm ds}$ was introduced as
\begin{equation}
A_{\rm ds}  \equiv (1+w) \, \xi \, .
\label{eq:Ads_mcmc}
\end{equation}
This parametrisation allowed for the clearly defined $\Lambda$CDM-limit: $A_{\rm ds}=0, w_0=-1$. However, when $w(a)$ is a time-dependent function, $A_{\rm ds}$ also becomes time-dependent. We then must choose a specific value of $a$ at which to define $A_{\rm ds} \equiv A_{\rm ds}(a_{\rm ref})$. In our analyses we will choose the pivot redshift of the main dataset under consideration, in this case DES Y3 + DESI DR2, to define $A_{\rm ds, piv}\equiv A_{\rm ds}(a_{\rm piv})$ with $a_{\rm piv}\approx 0.72$ ($z_{\rm piv} \approx 0.4$). This value was computed from the CPL-chain (first column in \autoref{tab:constraints}) following the same procedure outlined in~\citet{DES:2018ufa}. Additionally, we provide constraints on $A_{\rm ds}$, which in this works denotes $A_{\rm ds}(a=1)$. \refr{Finally, we point out that for $A_{\rm ds} = 0$ we get exactly CPL growth since the friction term in the Euler equation vanishes. Equivalently, away from the special case of $w(a_{\rm piv}) = -1$,  $A_{\rm ds, piv} = 0$ also gives CPL growth.}

Moving to nonlinear scales, the scale-independent linear suppression becomes scale dependent~\citep{Baldi:2014ica,Baldi:2016zom,Bose:2017jjx,Carrilho:2021otr}. To model these effects, we adopt the halo model reaction formalism~\citep{Cataneo:2018otr}. This framework models corrections to the nonlinear power spectrum through a halo model-based reaction function~\citep[we refer the reader to][for details]{Cataneo:2018otr,Bose:2021otr,Carrilho:2021otr}. The full nonlinear matter power spectrum is then given by
\begin{equation}
    P_{\rm NL}(k;a) = B(k;a) \, P_{\rm NL}^{\rm \Lambda CDM}(k;a) \, , 
    \label{eq:pnl}
\end{equation}
where the nonlinear boost is given by
\begin{equation}
    B(k;a) \equiv  \frac{ \mathcal{R}(k,a) \, P^{\rm pseudo}(k,a)}{P^{\rm \Lambda CDM}(k,a)}\, .
    \label{eq:boost}
\end{equation}
The reaction function $\mathcal{R}$, is computed using the halo model reaction code {\tt ReACT}~\citep{Bose:2020otr,Bose:2022vwi} and includes the effects of massive neutrinos and momentum exchange between dark matter and dark energy. The pseudo power spectrum is a nonlinear \lcdm{} spectrum prediction where the initial conditions have been tuned such that the linear clustering at the target redshift matches the beyond-\lcdm{} cosmology's linear clustering.  Given the scale-independent modification of the linear spectrum in DS, this amounts to changing the amplitude of the input \lcdm{} linear spectrum.

The \lcdm{} nonlinear ($P^{\rm \Lambda CDM}$) and pseudo ($P^{\rm pseudo}$) power spectra are computed in this work using an {\tt HMCode2020}~\citep[\href{https://github.com/alexander-mead/HMcode}{\faicon{github}}]{Mead:2020vgs} emulator (\href{https://github.com/MariaTsedrik/HMcode2020Emu/tree/main}{\faicon{github}}). The pseudo power spectrum is computed simply by specifying the DS linear power spectrum to {\tt HMCode2020}. We construct a neural network emulator for $B(k;a)$ to calculate the nonlinear power spectrum efficiently. Details are given in the next section.

\section{Methods and data}
\label{sec:data_and_methods} 

\subsection{Nonlinear boost emulation}

\begin{table}
\centering
\caption{Flat priors for DS-boost emulator. Massive neutrinos are included via the halo model reaction. Baryonic feedback is not included. In the last rows we include additional priors used in the actual analyses.}
\begin{tabular}{ | c | c | c | } \hline 
 {\bf Parameter} & {\bf  Prior} \\ \hline 
$\Omega_{\rm m,0}$ & [0.22, 0.37]  \\ 
$\Omega_{\rm b,0}$ & [0.03, 0.08]  \\ 
$\Omega_{\rm \nu,0}$ & [0.00015, 0.00317]  \\ 
$h$ & [0.63, 0.84]  \\ 
$n_s$ & [0.8, 1.1]  \\ 
$A_s 10^9$ & [1.7, 2.5]  \\ \hline
$w_0$ & [-1.3, -0.5] \\ 
$w_a$ & [-2, 0.5] \\ 
$\xi$ [bn/GeV] & [0,150] \\ \hline
$z$ & [0, 2.5] \\ \hline \hline
$A_{\rm ds, piv}$ [bn/GeV] & [-30, 30] \\ 
$\omega_{\rm b}$ & $\mathcal{N}(0.02268, 0.00038)$ \\ 
$M_\nu$ [eV] & 0.06 \\ 
$\tau$ & $\mathcal{N}(0.054, 0.0074)$ \\ \hline
\end{tabular}
\label{tab:priors}
\vspace{2mm}
\end{table}

\begin{figure}
    \includegraphics[width=1\columnwidth]{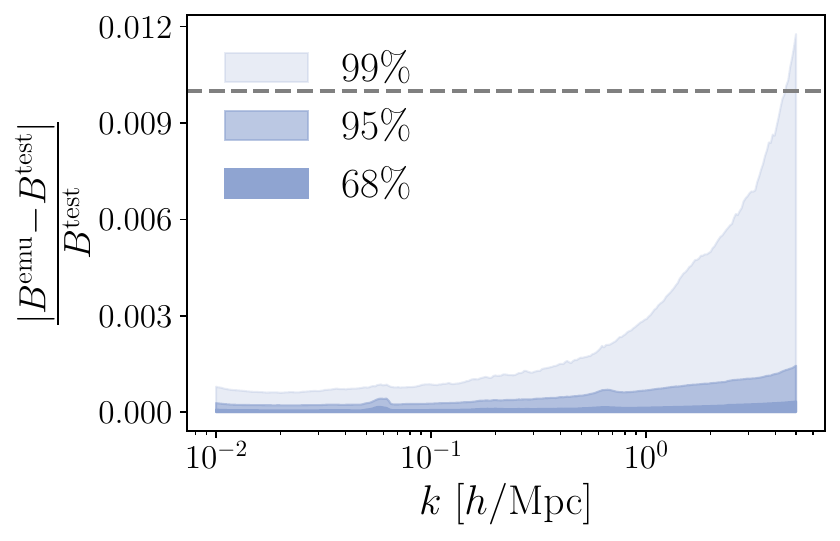}
    \caption{Quantiles of the relative accuracy of our DS emulator on $6 \cdot 10^4$ testing samples. The dashed grey line denotes the 1\% accuracy.}
    \label{fig:validation}
\end{figure}

To perform fast cosmological analyses, particularly for weak lensing measurements which require specification of the nonlinear power spectrum, we construct a neural network emulator for \autoref{eq:boost}. This is done using the {\tt Cosmopower} platform~\citep{SpurioMancini:2021ppk}. We keep the same architecture and values of hyperparameters from the original paper.

For the training and validation set boost data, we sample a latin hypercube defined by the parameters and priors given in \autoref{tab:priors}. Within this
range we generate $1.4 \cdot 10^{6}$ spectra 
for the training set, with an additional $6 \cdot 10^{4}$ generated for testing. We produce boosts in the range $k \in [10^{-2}, 5]~h$/Mpc. We restrict the redshift range and DS parameters based on stability criteria of {\tt ReACT}. In particular, very low values of $\sigma_8$ can result in issues within the spherical collapse part of the code. This has been partly remedied with a \href{https://github.com/nebblu/ACTio-ReACTio/tree/nebblu-patch-scol_initial_guess}{recent patch} to the code. The emulator can be found here \href{https://github.com/nebblu/ReACT-emus/tree/main/emulators/Dark%20Scattering}{\faicon{github}}.

In \autoref{fig:validation} we show the performance of the emulator on the test data set, indicating that further training data is not needed for our analyses. We note that \autoref{eq:boost} is at best $~1\%$ accurate when compared to simulations at $k\in[0.01,2]~h/{\rm Mpc}$ and for $z\in[0,1]$ for modifications charactertised by $\xi \leq 50~[{\rm bn/GeV}]$~\citep{Carrilho:2021otr} . We have further validated our emulator in the case of $w_a=0$ against the emulators produced in \cite{Carrion:2024itc} through direct comparison, achieving less than $0.3\%$ agreement for the entire $k-$range for specific cosmologies. Lastly, a final check was performed by analysing the DES Y3 data for $w_a=0$ and comparing against the results of~\cite{Tsedrik:2025cwc}, with high agreement of the posteriors obtained. For the latter test, a disagreement in the constraint on $A_{\rm ds}$ (see \autoref{eq:Ads}) was found but this was a function of the prior adopted on $\xi$. Another difference is that the emulator used in \cite{Tsedrik:2025cwc} was trained in $A_{\rm ds}$ and not $\xi$ allowing the exploration of the posterior distribution close to $w\approx-1$ but with $\xi>150$ bn/GeV. This being said, both results were prior dominated.

\subsection{Data sets}

In our analyses we consider the following data sets:\\
{\bf DES}: We use the publicly available measurements of the three two-point correlation functions ($3\times2$pt) from the third data release of the Dark Energy Survey~\citep{DES:2021wwk, DES:2022ccp}. These include cosmic shear, galaxy clustering, and galaxy–galaxy lensing, measured from sources in four redshift bins and lenses from the first four redshift bins of the MagLim sample~\citep{DES:2020ajx}. The corresponding covariance matrix is computed analytically following \citet{DES:2020ypx}, where the methodology is also validated. We adopt the same scale cuts as in the DES Y3 \lcdm{} baseline analysis. Our halo model reaction approach to nonlinearities ensures we have sufficient accuracy to probe these scales. We neglect baryonic feedback effects for reasons discussed in \cite{DES:2021rex}. \\
{\bf DESI}: We consider all tracers from DESI DR2 BAO \citep{DESI:2025zgx}.\\
{\bf CMB}: We include \textit{Planck 2018} TT, TE, EE temperature and polarization with \texttt{Planck-lite} likelihood \citep{Prince:2019hse, Planck:2019nip} including 2 low-$\ell$ in TT. \refr{This excludes the CMB-lensing potential, but the spectra intrinsically includes lensing information}.  For cosmological analyses, we apply the Planck-based Gaussian prior on the optical depth of reionization, $\tau$ \citep{planck2018cosmo,Balkenhol:2024obe}. We validate our choice of $\tau$-prior by comparing to a Cobaya chain with full Planck likelihood for DESI+SN+CMB -- both in CPL and DS our constraints in $\tau$ are not prior dominated. In analyses with CMB we do not impose a BBN prior on $\omega_\mathrm{b}$ as it is tightly constrained by the CMB information. \\
{\bf SN}: We add the full DES Y5 type Ia supernovae sample \citep{DES:2024jxu}. This includes 1635 SNe from $z>0.1$ and 194 SNe at $z<0.1$.

For our analyses we make use of the official DES pipeline, \texttt{CosmoSIS} 
(\href{https://github.com/joezuntz/cosmosis-standard-library}{\faicon{github}}), in which we include two new modules, one which calculates and applies the nonlinear and linear boost factors to the nonlinear and linear power spectra respectively, and one which calculates the linear growth factor and rate. The latter is done using the {\tt MGrowth}~(\href{https://github.com/MariaTsedrik/MGrowth}{\faicon{github}}) code. The boosts are applied to \texttt{CAMB}~\citep{camb} produced linear and {\tt HMCode2020}-based nonlinear spectra. This makes our setup identical to \cite{Tsedrik:2025cwc} except we apply a boost rather than emulating the nonlinear and linear spectra directly. CMB observables are computed with the modified version of \texttt{CLASS} for DS~\citep{alessa2025parameterized}. For chains without CMB we use \texttt{polychord} \citep{Handley:2015fda} with ``publication quality'' settings from \citet{DES:2022ykc}, for chains with CMB we use \texttt{Nautilus} \citep{Lange:2023ydq} with 12000 live points. We changed the sampler due to more efficient convergence of the latter. We cross-checked both samplers in several configurations: contours and $\ln{\mathcal{Z}}$ were in agreement, while the uncertainty on  $\ln{\mathcal{Z}}$ with \texttt{Nautilus} was $\mathcal{O}(10)$ tighter than with \texttt{polychord}.


\begin{figure*}
\centering
    \includegraphics[width=0.8\textwidth]{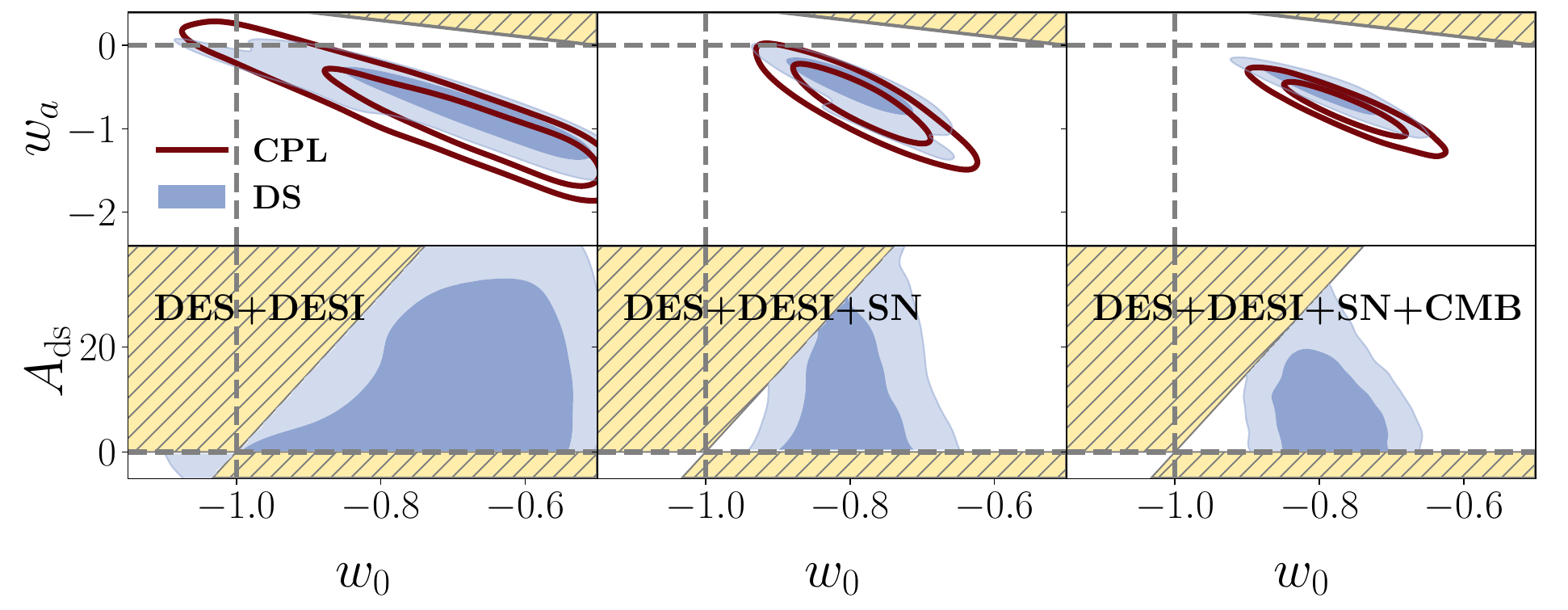}
    \caption{Dark energy results for different data selections, specified in the legend of the lower panels. CPL constraints are shown in dark red, DS constraints are shown in blue. All contours shown contain 68\% and 95\% of the posterior probability. Dashed yellow regions denote priors in DS: $w_0+w_a<-1/2$ for the existence of a growing solution of the linearised growth equation and $0<\xi<150$ [bn/GeV].}
    \label{fig:w0waAds}
\end{figure*}

\section{Results}
\label{sec:results}

\begin{figure}[t]
\centering
    \includegraphics[width=0.6\columnwidth]{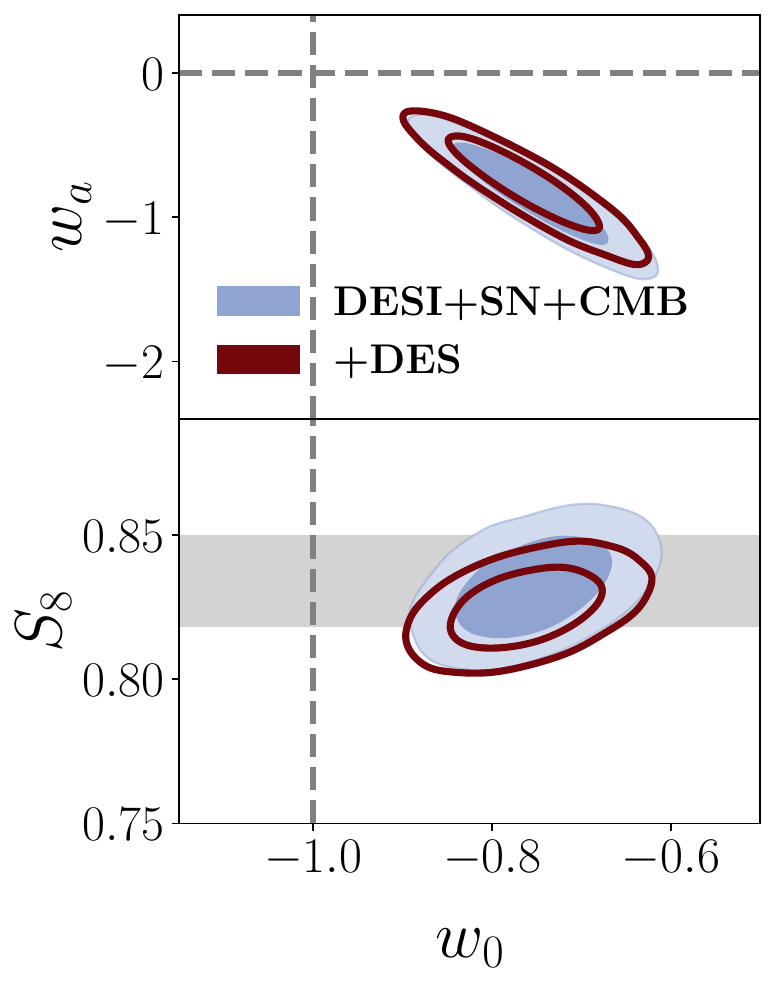}
    \caption{CPL: constraints without (in blue) and with (in dark red) photometric information from DES. Grey dashed lines denote \lcdm{} limit in equation-of-state parameters. Grey stripe denotes constraints from \textit{Planck 2018} without lensing for $\Lambda$CDM \citep{planck2018cosmo}.}
    \label{fig:cpl+spectro+photo}
\end{figure}

We present our full set of results in \autoref{tab:constraints} and \autoref{tab:constraints_all} (see Appendix). In all data combinations we find $A_{\rm ds,piv}$ to be consistent with zero within $1\sigma$. The exception is the DESI+SN+CMB combination, which shows a $\sim$$2\sigma$ preference for a non-zero interaction. In this case, the posterior hits the prior boundary because the interaction is unconstrained by background probes and the CMB data set is only very mildly constraining \citep{Pourtsidou:2016ico}. This indicates projection effects caused by degeneracies between the standard cosmological and extended parameters \citep{gomezvalent2022,Hadzhiyska:2023wae, Carrilho:2022mon}. All combinations also show a preference for $(w_0,w_a)$ in the region favoured by DESI, as expected.

The inclusion of CMB data drives $S_8$ to slightly higher values, as has been noted previously in the literature~\citep[see e.g.][]{DES:2022ccp}. Furthermore, the DS model tends to enhance structure growth, pushing $S_8$ to systematically higher values than in the CPL case. This behaviour is expected given the region of $(w_0,w_a)$ space preferred by DESI data (see \autoref{fig:linear-growth}). Our CPL results are in good agreement with the DESI results presented in \cite{DESI:2025zgx} for the DES+DESI+SN and DESI+SN+CMB combinations. The former shows a slight discrepancy in $w_a$, which we attribute to differences in priors on cosmological parameters between the official DES Y3 analysis \citep{DES:2022ccp} adapted by DESI and our choices summarised in \autoref{tab:priors}. An additional, yet less significant, difference between the DESI and our results is the choice of CMB likelihoods. 

We further compare DS and CPL in the $w_0–w_a$ and $w_0–A_{\rm ds}$ planes in \autoref{fig:w0waAds}. The DS contours are marginally shifted toward higher values of $w_a$ across all combinations with the DES data. Higher values of $w_a$ reduce the enhancement of structure growth induced by the interaction in this region of $(w_0,w_a)$ space (see \autoref{fig:linear-growth}). Since DES data prefers slightly lower values of $S_8$, this leads to a mild shift in the DS contours. The extent of the shift is restricted by SN data which prefers lower values of $w_a$. We note here that recalibration of the DES Y5 SN may weaken its preference for more negative values of $w_a$~\citep{DES:2025sig}.

\begin{figure}[t]
\centering
    \includegraphics[width=0.6\columnwidth]{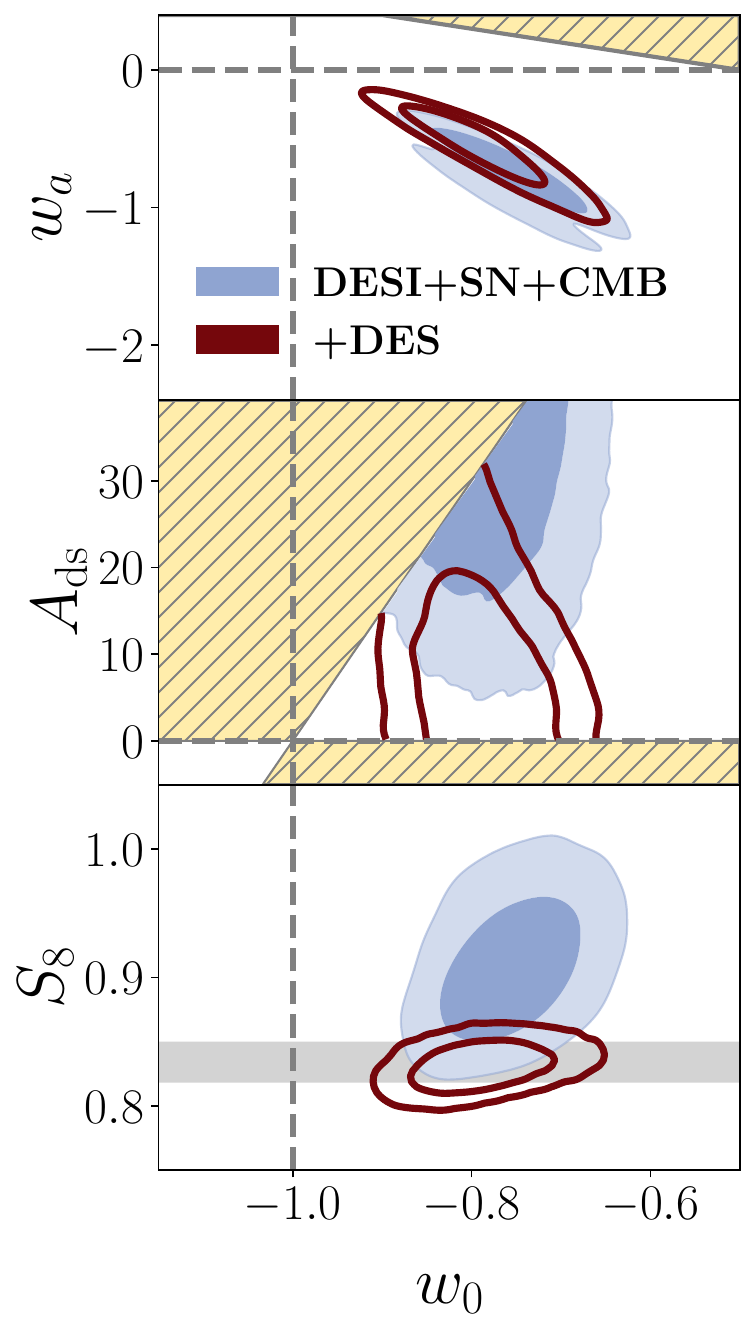}
    \caption{Same as \autoref{fig:cpl+spectro+photo} for DS. Dashed yellow regions denote prior constraints same as in \autoref{fig:w0waAds}.
    }
    \label{fig:spectro+photo}
\end{figure}

This behaviour is further illustrated in \autoref{fig:cpl+spectro+photo} and \autoref{fig:spectro+photo}, which show the impact of DES on the $w_0$, $w_a$, and $S_8$ constraints for CPL and DS, respectively. In the CPL case, we find that DES does not significantly improve constraints on $w_0$ or $w_a$ (or $S_8$) consistent with the findings of \cite{DES:2022ccp}. In DES, the preference for $S_8$ values mildly lower than the ones extrapolated from DESI+SN+CMB appears to tighten constraints on $S_8$ in the full combination of probes and lead to a slight improvement in $(w_0w_a)$.  In contrast, for the DS model, the inclusion of DES removes the $\sim$$2\sigma$ preference for non-zero $A_{\rm ds}$ seen in the background+CMB combination, and consequently tightens the $S_8$ constraint towards lower values.  This is also reflected in the improvement of the Figure-of-Merit (FoM) when adding the DES data to the DESI+CMB+SN combination. 
In CPL we find  \refr{FoM=173.31 $\rightarrow$ FoM=194.76} which constitutes a \refr{12.4\%} improvement by including DES. For DS we have \refr{FoM=180.13 $\rightarrow$ FoM= 224.46} which is a larger \refr{24.6\%} improvement, highlighting how the model requires structure formation data in order to be constrained. 

\refr{Regarding \autoref{fig:spectro+photo}, we note that the primary CMB spectra are lensed and so include a degree of late-time structure formation information. This can result in the slight preference for $A_{\rm ds}>0$ in the DESI+SN+CMB combination which naturally mimics $A_{\rm L}>1$, the well-known CMB lensing amplitude anomaly~\citep[see for example][]{CosmoVerse:2025txj}. Furthermore, the DS model will induce an enhanced Integrated-Sachs-Wolfe effect~\citep{Pourtsidou:2016ico} at large scales, but also a modulation of the smoothing of the peaks at small scales. Finally, an extended parameter space and non-Gaussian posteriors with weakly constrained parameters are at risk of inducing projection effects, resulting in apparent shifts of the marginalised posteriors~\citep{Tsedrik:2025hmj}.}

As with previous analyses~\citep[see for example][]{Tsedrik:2025cwc} we find no signs of interaction nor of the $S_8$ tension. \refr{By $S_8$ tension we mean a discrepancy between the low-$z$ clustering amplitude derived from the high-$z$ CMB measurements in $\Lambda$CDM and from the low-$z$ measurements (BAO, SN and $3\times2$pt) in CPL and DS. See \autoref{fig:S8argument} of Appendix for an illustration of this point and constraints from \autoref{tab:constraints}. Once the late-time freedom of CPL or DS is introduced, however, the value of $S_8$ inferred from the CMB alone becomes significantly broadened due to degeneracies with the dark-energy parameters and the expansion rate. As a consequence, the CMB-derived $S_8$ constraints in CPL and DS are not in meaningful tension with the low-redshift measurements in these models.}

\begin{figure}[t]
\centering
    \includegraphics[width=0.7\columnwidth]{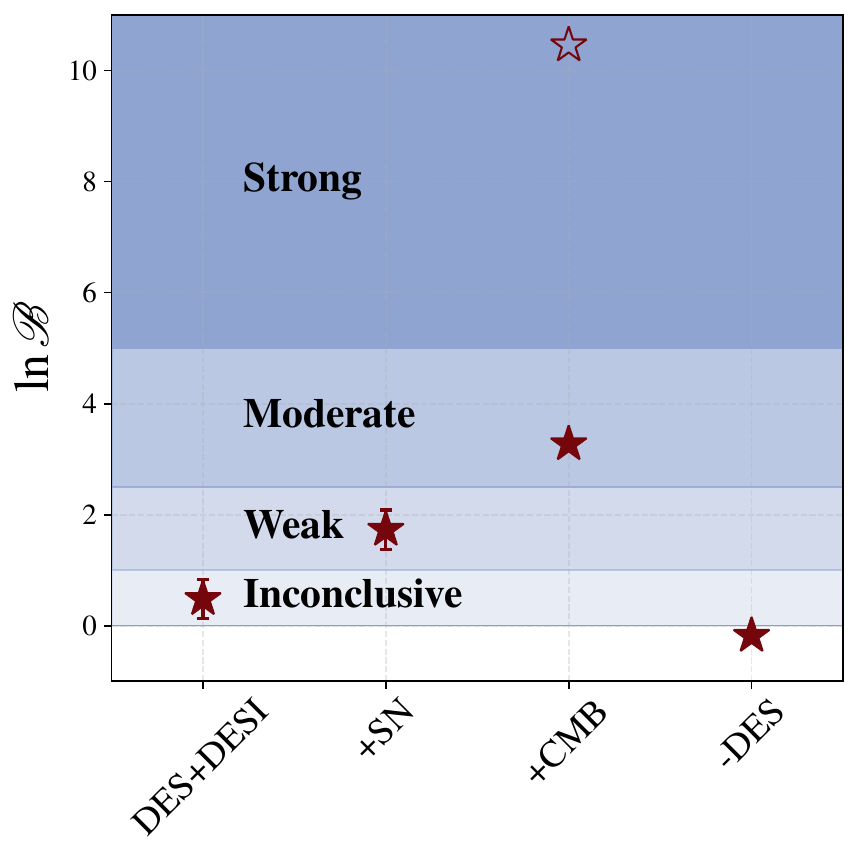}
    \caption{The Bayes factor: $\ln{\mathcal{B}}=\ln\mathcal{Z}_{\rm CPL}-\ln\mathcal{Z}_{\rm DS}$. Errorbars correspond to 68\% c.l. provided by the samplers. The shaded regions denote the Jeffreys' scale thresholds for evidence interpretation, i.e. significance of CPL being better supported by data than DS:  $0<\ln{\mathcal{B}}<1$ (inconclusive), $1<\ln{\mathcal{B}}<2.5$ (weak), $2.5<\ln{\mathcal{B}}<5$ (moderate), and $\ln{\mathcal{B}}>5$ (strong). The filled stars denote DS with the evolving equation-of-state, while the hollow star denotes DS with the constant equation-of-state.}
    \label{fig:Bayes-factor}
\end{figure}

Finally, in \autoref{fig:Bayes-factor} we present the Bayes factor comparing CPL and DS across different data combinations. Shaded regions represent the Jeffreys' scale thresholds for the Bayes factor interpretation \citep{Trotta:2008qt}. These results show that DS does not alleviate the mild preference for lower $S_8$ values from DES when CMB data are included, as the interaction enhances structure growth in the region of $(w_0,w_a)$ space favoured by DESI. Adding SN+CMB data to DES+DESI increases the preference for CPL over DS from negligible to moderate, while removing DES reduces this preference back to negligible. 

For the full data combination we repeat the analysis in DS with the constant equation-of-state. In agreement with \citet{DESI:2025zgx}, for a $w_0$CDM background and this combination of probes we obtain $w_0 \approx -0.96$, $\Omega_{\rm m}$ lower and $h$ higher than in $w_0w_a$CDM. In this scenario, for $w_0>-1$, any non-zero interaction between dark matter and dark energy leads to suppression of LSS growth at late times. Hence, we obtain a low value of $S_8$ not supported by the data. Overall, CPL shows strong evidence over DS with constant $w(a)=w_0$. The corresponding posterior distributions in all cosmological parameters for the full combination of probes is shown in \autoref{fig:posterior} of Appendix.


\section{Conclusions}
\label{sec:conclusions}
In this paper we combine the 3$\times$2pt correlation functions of DES Y3, and the BAO measurements of DESI DR2, together with DES Y5 SN and Planck CMB, in order to constrain the DS model -- a momentum exchange interacting dark energy model. For reference we compare our results for each data combination with constraints from the CPL model. Both DS and CPL have the same background evolution, with the equation-of-state for dark energy given by $w=w_0+w_a(1-a)$. DS includes an additional parameter, $\Ads$ or $\xi$, which controls the strength of the coupling between dark matter and dark energy. This parameter does not depend on the background observables, SN and BAO, and only weakly modifies low-$\ell$ and lensing in CMB. However, the non-zero interaction parameter enhances or suppresses growth of structures in a way, which depends on values of the equation-of-state parameters. Hence, cosmological probes, which contain information on the time-evolution of LSS growth, are essential to constrain DS and differentiate it from CPL.

From the full combination of cosmological probes we obtain the following constraints: $w_0=-0.76\pm 0.06$ and $w_a=-0.77^{+0.23}_{-0.20}$ for CPL, $w_0=-0.79^{+0.05}_{-0.06}$, $w_a=-0.56^{+0.24}_{-0.15}$, and $\Ads=9.8^{+2.8}_{-9.5}$ bn/GeV for DS. There is no detection of interaction, with $\Ads$ being consistent with zero. Moreover, the equation-of-state parameter space preferred by DESI+SN+CMB, leads to enhancement of structure growth for non-zero $\Ads$. This dependence between $(w_0,w_a)$  and $\Ads$ drifts the constraints in $w_a$ towards less negative values than in CPL in an attempt to cross the enhancement/suppression transition we show in \autoref{fig:linear-growth}. However, the combined signal from DESI+SN favours a restricted region in $(w_0,w_a)$ space. We note that the DES Y5 SN sample we use in our analysis has been recently re-calibrated \citep{DES:2025sig}. This might result in slightly less negative values of $w_a$ and a $(w_0,w_a)$  combination that allows a potential scattering signal. Hence, we treat our constraints from the chosen data combination as rather pessimistic for DS detection and leave the investigation with the re-calibrated sample for future work.   

We quantify the role of LSS-growth data, by assessing the FoM in $(w_0,w_a)$  without and with addition of the 3$\times$2pt information from DES to DESI+SN+CMB. For CPL, the improvement in constraints with DES is not significant: \refr{$\sim$12\%}. This is similar to the addition of the full-shape information from the spectroscopic measurements of galaxy distributions~\citep{Chudaykin:2025lww}. However, for DS this number increases up to \refr{$\sim$25\%} due to the constraining power in $\Ads$ through the well-measured amplitude, $S_8$.

Both in CPL and DS, we find no evidence for the $S_8$ tension. This is consistent with our previous findings in \citet{Tsedrik:2025cwc}. The constraints on this late-time amplitude agree well with the predictions from CMB in \lcdm{}. For CPL, the  $(w_0,w_a)$ values remain on the already established region that DESI+SN+CMB prefer, which falls on a line in $(w_0,w_a)$  space that gives no relative change to growth over \lcdm{}. For DS, this growth-enhancement/suppression line is a function of the scattering amplitude, with the preferred region of $(w_0,w_a)$  values lying in the enhancement regime. Consequently, DS would be preferred if the DES measurement of $S_8$ would be higher than the one derived from the CMB constraints. This is the opposite of the observed trend in other lensing data sets~\citep[e.g.~Hyper Suprime-Cam][]{Dalal:2023olq,Sugiyama:2023fzm,Miyatake:2023njf,Li:2023tui}.

\refr{To further clarify, the CMB alone provides only weak constraints on $S_8$ in DS and CPL models as late-time modifications can significantly change its value. The late-time clustering amplitude is mainly determined by the DES $3\times2$pt measurements (see Fig.~\ref{fig:spectro+photo} for example). As shown in Fig.~\ref{fig:S8argument} of Appendix, the values of $S_8$ preferred by the low-redshift probes are consistent with the broader range allowed by the CMB in CPL and DS, and thus no statistically significant tension is present.}

While providing consistent constraints on shared cosmological parameters, the preference for CPL over DS becomes clear by evaluation of the Bayes factor. In all data combinations with DES, CPL is preferred by data, consistent with the argumentation in the previous paragraph. Additionally, we perform an analysis for the constant equation-of-state in DS with the full combination of probes -- CPL remains  better supported by the data in this case too.

There are alternative interacting dark energy models, such as the one discussed in \citet{Li:2025ula}. There, dark energy and dark matter exchange both momentum and energy, while the equation-of-state for dark energy is fixed to $w=-1$. The flow of energy between dark matter and dark energy reverses direction in the late-time Universe. The authors claim that this model explains DESI+SN+CMB measurements as well as CPL. However, the parameter space in the ``sign-reversal'' model preferred by this combination of data ultimately leads to low values of $S_8$ -- hence less preferable by DES $3\times2$pt baseline analysis. 
This highlights an importance of structure growth information in constraining interacting dark energy models. With the new generation of photometric surveys like Euclid and Rubin, we will have more certainty in whether and how dark energy and dark matter interact.

\section*{Acknowledgements}

The authors are grateful to the DESI and DES collaborations for making their data publicly available. The authors thank Pedro Carrilho for valuable comments, and Durrah Alessa for a discussion on the PPF framework with interacting dark energy. BB is supported by a UK Research and Innovation Stephen Hawking Fellowship (EP/W005654/2). MT's research is supported by grant ST/Y000986/1. For the purpose of open access, the author has applied a Creative Commons Attribution (CC BY) license to any Author Accepted Manuscript version arising from this submission.

\section*{Data Availability}

Links and references with sources of the data and analysis pipelines are provided in the main text. 



\bibliographystyle{mnras}
\bibliography{refs} 



\appendix

\begin{table*}[h]
  \small
	\centering
	\caption{Mean values and 68\% c.l. values for CPL and DS with fixed
    one massive neutrino with $M_{\nu}=0.06~{\rm eV}$ 
    for the different combination of probes considered in this work. The values are computed with \texttt{getdist} \citep[\href{https://github.com/cmbant/getdist}{\faicon{github}}]{Lewis:2019xzd} with imposed ranges from \autoref{tab:priors}. We show the maximum \textit{a posteriori} values from the chains in parentheses. In the second vertical block, we include derived constraints on $\Ads$ (\autoref{eq:Ads_mcmc} at $z=0$), $\sigma_8$, $S_8=\sigma_8\sqrt{\Omega_{\rm m}/0.3}$, $\ln{\left(10^{10} A_\mathrm{s} \right)}$. We report the number of data-points $N_{\rm data}$, the number of parameters $N_{\rm par}$ varied in the MCMC, 
    and log-evidence $\ln{\mathcal{Z}}$ from the sampler. Unconstrained parameters are denoted by a dash-line. The BBN prior is applied, sampled with \texttt{polychord}. 
    }
\label{tab:constraints}
	\begin{tabular}{c | cc | cc }	
		 & \multicolumn{2}{c|}{{\bf DES+DESI}} & \multicolumn{2}{c}{{\bf DES+DESI+SN}}  \\ 
         $N_{\rm data}$ & \multicolumn{2}{c|}{462+13} & \multicolumn{2}{c}{462+13+1829}   \\ 
         \hline 
		& CPL & DS & CPL & DS   \\
		\hline
		$\Omm$ & $0.330^{+0.019}_{-0.012}(0.337)$  & $0.327^{+0.016}_{-0.012}(0.357)$ & $0.321\pm 0.009(0.313)$ & $0.317\pm 0.008(0.323)$ \\
        $\omega_{\rm b}\cdot 10^{2}$ & $2.27\pm 0.04 (2.23)$ & $2.27\pm 0.04(2.26)$ & $2.27\pm 0.04(2.27)$ & $2.27\pm 0.04(2.26)$ \\ 
        $w_0$ & $> -0.740(-0.643)$ & $-0.688^{+0.18}_{-0.048}(-0.5)$  & $-0.782\pm 0.063(-0.795)$ & $-0.792^{+0.051}_{-0.059}(-0.764)$  \\
        $w_a$ & $-0.93^{+0.33}_{-0.58}(-1.37)$ & $-0.78^{+0.32}_{-0.41}(-1.4)$  & $-0.71^{+0.33}_{-0.29}(-0.66)$ & $-0.54^{+0.30}_{-0.18}(-0.78)$ \\
        $\Adsp$ & 0 & $5.4^{+2.5}_{-5.8}(0.9)$ & 0 & $4.0^{+1.9}_{-4.7}(0.5)$  \\
        $A_{\rm s}\cdot 10^{9}$& $- (1.741)$  & $-(1.804)$  & $-(1.993)$ & $-(1.804)$ \\
        $h$& $0.664^{+0.012}_{-0.017}(0.673)$ & $0.655^{+0.009}_{-0.017}(0.672)$  & $0.674\pm 0.009(0.664)$ & $0.667\pm 0.009(0.676)$  \\
        $\ns$ & $0.962^{+0.039}_{-0.045}(0.951)$ & $0.988\pm 0.040(0.942)$ & $0.962\pm 0.043(1.004)$  & $0.989^{+0.043}_{-0.038}(0.946)$\\

        \hline
         $\Ads$ & 0 & $17.0^{+6.7}_{-17}(4.4)$ & 0 & $12.2^{+4.5}_{-12}(7.9)$\\
        $\sigma_8$ & $0.777^{+0.018}_{-0.026}(0.773)$ & $0.781^{+0.019}_{-0.022}(0.740)$ & $0.785^{+0.016}_{-0.021}(0.791)$  & $0.790\pm 0.018(0.788)$\\
        $S_8$ & $0.814 \pm 0.015(0.819)$ & $0.816\pm 0.018(0.808)$  & $0.811\pm 0.015(0.808)$  & $0.812\pm 0.017(0.818)$\\
        $\ln{\left(10^{10}A_{\rm s}\right)}$ & $-(2.857)$ & $-(2.893)$ & $-(2.992)$  & $-(2.893)$\\
		\hline
        $N_{\rm par}$ & 29 & 30 & 29 & 30\\
        $\ln{\mathcal{Z}}$ & $5740.1\pm 0.2$ & $5739.6 \pm 0.2$ &  $4915.3 \pm 0.2$ & $4913.5 \pm 0.2$\\
	\end{tabular}
\end{table*} 

\begin{table*}[h]
  \scriptsize
	\centering
	\caption{Same as above but for data combinations with CMB. The BBN prior is not applied, sampled with \texttt{Nautilus}. Additionally considered is DS with constant equation-of-state, i.e. $w_a=0$ and $\Adsp=\Ads$.}
\label{tab:constraints_all}
	\begin{tabular}{c | cc | ccc }	
    & \multicolumn{2}{c|}{{\bf DESI+SN+CMB}} & \multicolumn{3}{c}{{\bf DES+DESI+SN+CMB}} \\ 
		$N_{\rm data}$ & \multicolumn{2}{c|}{13+1829+615} & \multicolumn{3}{c}{462+13+1829+615}  \\ 
         \hline 
		& CPL & DS & CPL & DS  & DS \refr{($w=const$)} \\
		\hline
		$\Omm$ & $0.319\pm 0.006(0.323)$ & $0.319\pm 0.006(0.318)$ &  $0.318\pm 0.005(0.316)$ & $0.318\pm 0.005(0.321)$ & $0.310\pm 0.005(0.312)$  \\
        $\omega_{\rm b} \cdot 10^2$ & $2.24\pm 0.01(2.22)$ & $2.25\pm 0.01(2.28)$ &$2.24\pm 0.01(2.26)$ & $2.25\pm 0.01(2.24)$ & $2.25\pm 0.01(2.24)$\\ 
        $w_0$ & $-0.753\pm 0.058(-0.758)$ & $-0.754\pm 0.052(-0.694)$ &$-0.762\pm 0.056(-0.821)$ & $-0.789^{+0.048}_{-0.056}(-0.735)$ & $-0.959^{+0.018}_{-0.020}(-0.953)$ \\
        $w_a$ & $-0.84^{+0.25}_{-0.22}(-0.85)$ & $-0.76^{+0.22}_{-0.16}(-0.93)$ &$-0.77^{+0.23}_{-0.20}(-0.51)$ & $-0.56^{+0.24}_{-0.15}(-0.71)$ & 0 \\
        $\Adsp$ & 0 & $-(6.3)$ &0 & $3.0^{+1.0}_{-3.1}(1.2)$ & $1.1^{+0.3}_{-1.1}(1.2)$\\
        $A_{\rm s}\cdot 10^{9}$& $2.109\pm 0.029(2.081)$ & $2.099\pm 0.030(2.087)$ & $2.098\pm 0.027(2.081)$ & $2.090\pm 0.028(2.057)$ & $2.112\pm 0.028(2.121)$\\
        $h$ & $0.668\pm 0.006(0.666)$ & $0.666^{+0.005}_{-0.006}(0.666)$ & $0.668\pm 0.006(0.669)$ & $0.666\pm 0.005(0.661)$ & $0.673\pm 0.005(0.670)$\\
        $\ns$ & $0.966\pm 0.004(0.967)$ & $0.968\pm 0.004(0.969)$ &  $0.968 \pm 0.004(0.968)$ & $0.970\pm 0.004(0.972)$ & $0.972\pm 0.003(0.975)$\\
        \hline
        $\Ads$ & 0 & $-(44.8)$ & 0 & $9.8^{+2.8}_{-9.5}(4.9)$ & $1.1^{+0.5}_{-1.1}(1.2)$\\
        $\sigma_8$ & $0.807\pm 0.010(0.808)$ & $0.881^{+0.032}_{-0.040}(0.925)$ & $0.801\pm 0.009(0.791)$ & $0.808^{+0.013}_{-0.014}(0.790)$ & $0.785\pm 0.011(0.784)$\\
           $S_8$ & $0.832\pm 0.012(0.839)$ & $0.909^{+0.034}_{-0.041}(0.953)$ & $0.825\pm 0.009(0.812)$ & $0.831\pm 0.014(0.818)$ & $0.797^{+0.010}_{-0.009}(0.799)$\\
        $\ln{10^{10}A_{\rm s}}$ & $3.049\pm 0.014(3.035)$ & $3.044\pm 0.014(3.038)$ & $3.044\pm 0.013(3.035)$ & $3.040\pm 0.013(3.024)$ & $3.050\pm 0.013(3.054)$\\
		\hline
        $N_{\rm par}$ & 8 & 9 & 30 & 31 & 30\\
        $\ln{\mathcal{Z}}$ & $-1140.659\pm0.006$ & $-1140.480\pm 0.006$ & $4612.566 \pm 0.006$ & $4609.294 \pm 0.006$ & $4602.111\pm0.006$\\
	\end{tabular}
\end{table*} 

\begin{figure*}[h]
\centering
    \includegraphics[width=\textwidth]{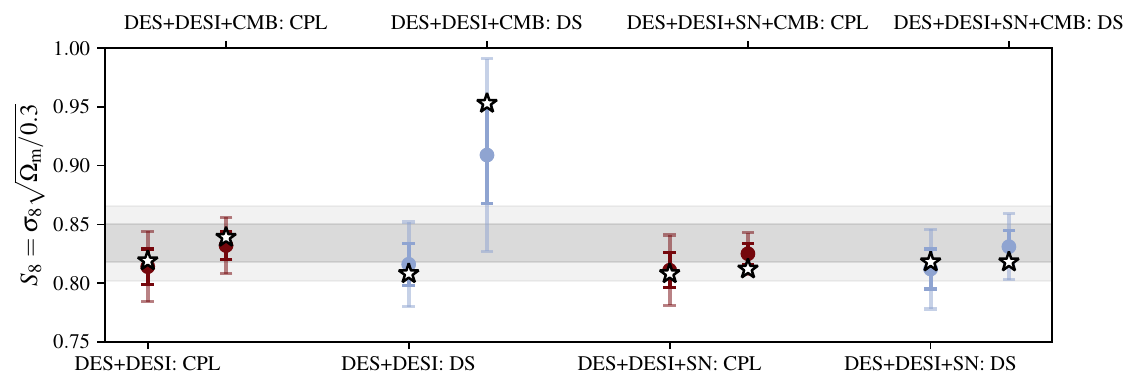}
    \caption{\refr{Marginalised constraints on $S_8$ in CPL and DS from the low-$z$ measurements (lower axis ticks) and their combination with CMB (upper axis ticks). The error bars show the 68\% marginalised posteriors (and 95\% in fainter colours), the stars denote the MAP values from the chains (see \autoref{tab:constraints}). The grey shaded lines correspond to constraints on $S_8$ from the high-$z$ CMB data with \textit{Planck 2018} without lensing for $\Lambda$CDM \citep{planck2018cosmo}. The constraints on $S_8$ agree within $2\sigma$ (including the projection dominated case of DES+DESI+CMB: DS). The MAP values with the SN information show consistent values across both models with or without CMB.}}
    \label{fig:S8argument}
\end{figure*}

\begin{figure*}[h]
\centering
    \includegraphics[width=\textwidth]{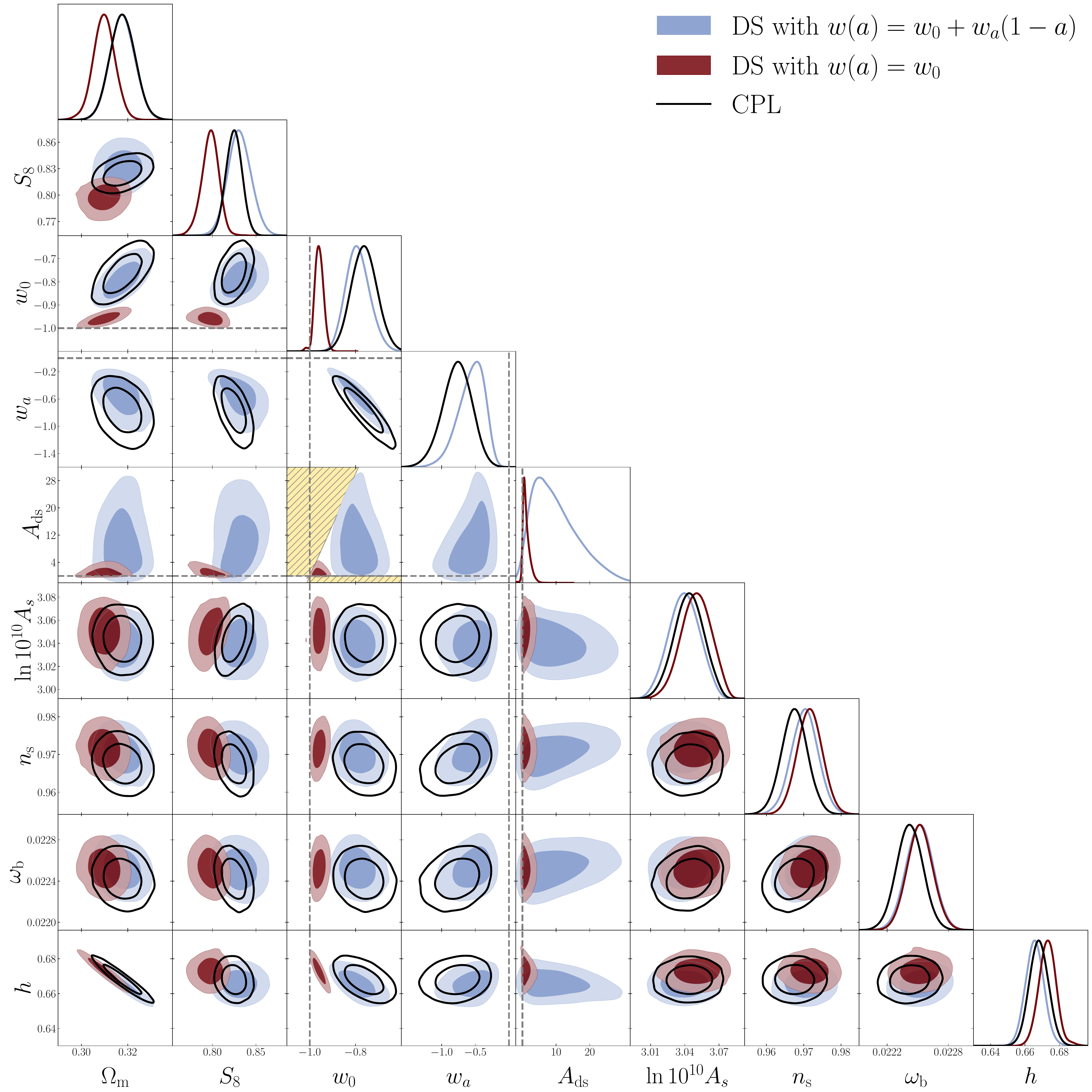}
    \caption{Dark energy results with the full combination of cosmological probes -- DES+DESI+SN+CMB -- for different models, specified in the legend. CPL constraints are shown in black, DS constraints with $(w_0w_a)$ are shown in blue and DS with the constant equation-of-state in dark red. All contours shown contain 68\% and 95\% of the posterior probability. Dashed yellow regions denote priors in DS: $w_0+w_a<-1/2$ for the existence of a growing solution of the linearised growth equation and $0<\xi<150$ [bn/GeV]. Dashed grey lines denote the \lcdm{} limit in $(w_0w_a)$ and $\Ads$.}
    \label{fig:posterior}
\end{figure*}

\label{lastpage}
\end{document}